\documentclass[conference]{IEEEtran}
\usepackage{graphicx} 
\usepackage{times}
\usepackage{mathtools}
\usepackage{amsmath}
\usepackage{amssymb}

\usepackage{algorithm}
\usepackage{algpseudocode}
\usepackage{caption}
\usepackage{url}
\usepackage{float} 
\usepackage{xcolor} 

\usepackage{orcidlink}

\date{}
\begin{document}

\title{Layered Normalized Min-Sum Decoding with\\ Bit Flipping for FDPC Codes}
\author{Niloufar Hosseinzadeh}

\author{Niloufar~Hosseinzadeh\textsuperscript{\href{https://orcid.org/0009-0002-0404-1590}{\includegraphics[scale=0.06]{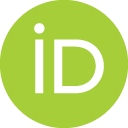}}},
Mohsen~Moradi\textsuperscript{\href{https://orcid.org/0000-0001-7026-0682}{\includegraphics[scale=0.06]{figs/ORCID}}},
and Hessam~Mahdavifar\textsuperscript{\href{https://orcid.org/0000-0001-9021-1992}{\includegraphics[scale=0.06]{figs/ORCID}}}\\
Department of Electrical and Computer Engineering, Northeastern University, Boston, MA 02115, USA \\ 
Email: hz93niloufar@gmail.com, \{m.moradi, h.mahdavifar\}@northeastern.edu\\
\thanks{The authors are with the Department of Electrical \& Computer Engineering, Northeastern University, Boston MA-02115, USA (e-mail: hz93niloufar@gmail.com, m.moradi@northeastern.edu, h.mahdavifar@northeastern.edu).}
\thanks{This work was supported by the National Science Foundation (NSF) under Grant CCF-2415440 and the Center for Ubiquitous Connectivity (CUbiC) under the JUMP 2.0 program.}
}

\maketitle
\begin{abstract}
Fair-density parity-check (FDPC) codes have been recently introduced demonstrating improved performance compared to low-density parity-check (LDPC) codes standardized in 5G systems particularly in high-rate regimes. 
In this paper, we introduce a layered normalized min-sum (LNMS) message-passing decoding algorithm for the FDPC codes. We also introduce a syndrome-guided bit flipping (SGBF) method to enhance the error-correction performance of our proposed decoder.
The LNMS decoder leverages conflict graph coloring for efficient layered scheduling, enabling faster convergence by grouping non-conflicting check nodes and updating variable nodes immediately after each layer. In the event of decoding failure, the SGBF method is activated, utilizing a novel reliability metric that combines log-likelihood ratio (LLR) magnitudes and syndrome-derived error counts to identify the least reliable bits.
A set of candidate sequences is then generated by performing single-bit flips at these positions, with each candidate re-decoded via LNMS. The optimal candidate is selected based on the minimum syndrome weight. Extensive simulation results demonstrate the superiority of the proposed decoder. Numerical simulations on FDPC$(256,192)$ code with a bit-flipping set size of $T = 128$ and a maximum of $5$ iterations demonstrate that the proposed decoder achieves approximately a $0.5\,\mathrm{dB}$ coding gain over standalone LNMS decoding at a frame error rate (FER) of $10^{-3}$, while providing coding gains of $0.75-1.5\,\mathrm{dB}$ over other state-of-the-art codes including polar codes and 5G-LDPC codes at the same length and rate and also under belief propagation decoding. 

\end{abstract}
\section{Introduction}
Low-density parity-check (LDPC) codes \cite{gallager2003low,mackay1999good} have become a key error-correction mechanism in communication systems offering near-capacity performance through iterative message-passing decoding algorithms such as belief propagation (BP) and its low-complexity approximations, including the normalized min-sum (NMS) algorithm. These codes are particularly well-suited for high-throughput applications and have been integrated into standards such as in 5G for data channels \cite{3GPP38212} and 802.11 Wi-Fi \cite{tsatsaragkos2017reconfigurable}, with ongoing consideration for 6G systems \cite{wang2023road}.

Recently, fair-density parity-check (FDPC) codes were introduced and have demonstrated superior error-correction performance compared to 5G-LDPC codes in high-rate regimes and various short-to-medium block lengths \cite{mahdavifar2024high}. They are also shown to achieve coding gains even with a small number of decoding iterations \cite{mahdavifar2024high, moradi2025high}. Furthermore, FDPC codes provide a mechanism to characterize the minimum distance and weight distribution of the codes \cite{mahdavifar2024high}, thereby enabling analytical characterization of the error floor regions, which are often very difficult to obtain for LDPC codes. 

The promising performance of FDPC codes has been shown under BP decoding with flooding scheduling, in which all check nodes update messages concurrently before variable nodes are updated.
This approach, while straightforward, may result in slower convergence due to the delayed propagation of information. To mitigate this issue in classical LDPC codes, layered scheduling has emerged as an effective alternative, partitioning check nodes into non-conflicting layers, for example, by using graph coloring techniques on the conflict graph \cite{raeisi2021edge}.
By updating variable nodes immediately after each layer, layered scheduling accelerates convergence. The integration of NMS into layered frameworks makes the overall decoding scheme suitable for practical implementations \cite{jayawickrama2022improved}.

A main optimization problem in layered decoding is to optimize the decoding schedule. Recently, reinforcement learning (RL) techniques have been utilized to address this problem for LDPC codes \cite{habib2021decoding, habib2023reinforcement}, as well as for polar codes \cite{moradi2025enhancing}. These learning-based methods demonstrate the potential of adaptive scheduling but often require significant training overhead and may not generalize across all code families. Also, BP decoders can still fail primarily due to trapping sets or residual errors \cite{rigby2018augmented}. To address these shortcomings, post-processing techniques can be used, including bit-flipping algorithms that leverage syndrome information and reliability metrics to identify and correct erroneous bits \cite{chang2010check}.
List-based approaches extend this by generating multiple candidates and selecting the best based on criteria like minimum syndrome weight, thereby improving reliability without excessive complexity \cite{mahdavifar2024high}.

Motivated by the aforementioned advances in BP decoding and building on the promising strengths of FDPC codes, in this paper, we propose and analyze a novel message-passing algorithm for FDPC codes. We introduce a hybrid decoder that combines layered normalized min-sum (LNMS) decoding with syndrome-guided bit-flipping (SGBF) list correction. The LNMS phase employs conflict graph coloring for efficient layered scheduling, resulting in fast convergence. In cases of decoding failure, the SGBF phase is activated, utilizing a novel reliability metric that integrates log-likelihood ratio (LLR) magnitudes with \textit{error counts} derived from the syndrome, i.e., how many unsatisfied parity checks each bit is involved in. Then the bits deemed \textit{unreliable} are identified and flipped. A set of candidates is generated by single-bit flips on the unreliable positions, each re-decoded via LNMS, and the optimal candidate is chosen based on syndrome weight. This hybrid strategy enhances error correction while maintaining manageable complexity.

Extensive simulation results are provided for various short-to-medium block length FDPC codes. For instance, simulations on the FDPC$(256,192)$ code over the binary-input additive white Gaussian noise (BI-AWGN) channel demonstrate the effectiveness of the proposed decoding algorithm. With a bit-flipping set size of $128$ and only $5$ iterations of the LNMS message-passing algorithm, the proposed decoder achieves approximately a $0.5$\,dB coding gain over the standalone LNMS at a frame error rate (FER) of $10^{-3}$. The overall scheme, i.e., FDPC code together with the proposed decoding algorithm, provides coding gains of $0.75-1.5\,\mathrm{dB}$ over other state-of-the-art codes including polar codes and 5G-LDPC codes at the same length and rate and also under BP decoding. Similar performance gains are observed for FDPC$(256, 164)$ and FDPC$(1024, 844)$ codes. 

The remainder of the paper is organized as follows. Section II provides preliminaries on FDPC codes. Section III presents the proposed decoder, detailing layered scheduling, LNMS decoding, and SGBF correction. Section IV discusses numerical results, and Section V concludes the paper.

\section{Preliminaries}
\noindent\textbf{Notation convention.} Throughout the paper, vectors and matrices are denoted in bold, and all operations are carried out over the binary field.

\noindent\textbf{FDPC codes.} These codes have been recently introduced in \cite{mahdavifar2024high} targeting high-rate applications. In the original construction, the base parity-check matrix $\mathbf{H}_b$ is a $2\sqrt{N} \times N$ matrix, where $N$ is the code block length and $N = t^2$ for some even integer $t \geq 2$. Each column of $\mathbf{H}_b$ has weight 2, and each row has weight $t$. The matrix $\mathbf{H}_b$ consists of all binary column-vectors of length $2t$ and Hamming weight 2, with the indices of the two non-zero entries differing by an odd number. The rank of $\mathbf{H}_b$ is $2t - 1$, and the minimum distance of the code defined by $\mathbf{H}_b$ is $4$. The order-$s$ FDPC code, for $s \geq 2$, is constructed by applying $s-1$ distinct permutations (e.g., random permutations) to the columns of $\mathbf{H}_b$ and stacking them on top of each other, resulting in a parity-check matrix $\mathbf{H}$ of dimension $2s\sqrt{N} \times N$. The code dimension is at least $N - 2s\sqrt{N} + s$.

\noindent\textbf{FDPC codes with efficient encoders.} In \cite{moradi2025high}, a slightly modified construction of FDPC codes is proposed to enable low-complexity encoding, while preserving the desirable properties of the original codes. For a given $t$, the columns of the base matrix $\mathbf{H}_b$ are rearranged to form $\mathbf{H}_{\text{base}-(t,1)}$ where again each column has exactly two ones. The rearrangement is done by grouping columns such that the first $2t-1$ columns have consecutive ones, the next $2t-3$ columns have two zeros between the ones, and so on, until the last column has $2t-2$ zeros between the ones in the first and last rows. To facilitate systematic encoding, the first $2t$ columns are modified to form a bidiagonal square matrix $\mathbf{A}$ of size $2t \times 2t$, resulting in the modified base matrix $\mathbf{H}_{\text{mbase}-(t,1)} = [\mathbf{A} | \mathbf{B}]$.

A certain form of order-3 FDPC codes is also introduced in \cite{moradi2025high} with $\mathbf{H}_{\text{base}-(t,2)}$, a $2t \times \frac{t(t+1)}{2}$ matrix aiming for a minimum distance of 6, by carefully choosing the gaps between the two ones in each column. The base matrix is generated using Algorithm~2 in \cite{moradi2025high}, which iteratively places ones with increasing gaps and optionally removes columns to achieve the desired minimum distance. The full parity-check matrix $\mathbf{H}$ is then constructed using Algorithm~1 in \cite{moradi2025high}, which applies a specified number of random column permutations denoted as \text{num\_per} to the submatrix starting from column $m_{\text{size}}+1$ (where $m_{\text{size}} = 2t \times (\text{num\_per} + 1)$), stacks them, and optionally punctures columns to achieve the desired $N \leq t^2$. The first $m_{\text{size}}$ columns are set to a bidiagonal form for encoding. The parity-check matrix \(\mathbf{H}\) used in this paper follows this structure with the following form \cite{moradi2025high}:

\[
\mathbf{H} =
\begin{bmatrix}
1 & 0 & 0 & \cdots & 0 & \mathbf{C} \\
1 & 1 & 0 & \cdots & 0 & \pi_1(\mathbf{C}) \\
0 & 1 & 1 & \cdots & 0 & \vdots \\
\vdots & \vdots & \vdots & \ddots & \vdots & \vdots \\
0 & 0 & 0 & \cdots & 1 & \pi_{\text{num\_per}}(\mathbf{C})
\end{bmatrix},
\]
where \(\mathbf{C}\) is the submatrix of the base matrix consisting of columns from \(m_{\text{size}} + 1\) to the end, and \(\pi_i(\mathbf{C})\) denotes the \(i\)-th random column permutation of \(\mathbf{C}\). If \(N < t^2\), columns are punctured from the data part (after column \(m_{\text{size}}\)) to achieve the desired block length.

\noindent\textbf{Encoding process.} For the systematic FDPC code \cite{moradi2025high} with a code block length of $N$ and a data length of $K$,
the parity vector $\mathbf{p}$ of length $M=N-K$ is derived from the message vector $\mathbf{m}$,
and the codeword is given by $\mathbf{c} = [\mathbf{p}, \mathbf{m}]$.
Specifically, the parity bits are computed from the data bits using
$
    s_i = \sum_{j=1}^{K} m_j H_{i, M+j},
$
and obtained as
\begin{equation}
    p_i =
    \begin{cases}
        s_1, & i = 1, \\
        s_i + p_{i-1}, & i > 1.
    \end{cases}
\end{equation}

\section{Proposed Decoder}

Our proposed decoder consists of an LNMS decoding algorithm, followed by an SGBF operation. In the following, we explain our algorithm in detail.

\subsection{Layered Scheduling via Conflict Graph Coloring}

In the conventional BP decoding algorithm, all check nodes process messages simultaneously (flooding), and due to the presence of short cycles, this can result in slower convergence.
In contrast, the LNMS message-passing algorithm groups the check nodes into layers by leveraging graph coloring on a conflict graph. In a conflict graph, two check nodes are considered adjacent if they share a variable node. Each layer consists of non-conflicting check nodes that can be processed in parallel. Variable nodes are updated immediately after processing each layer, which enables faster information propagation across iterations.

The $M \times N$ parity-check matrix \(\mathbf{H}\) defines a bipartite graph with \(M\) check nodes and \(N\) variable nodes. 
The conflict graph \(\mathcal{G} = (\mathcal{V}, \mathcal{E})\) models conflicts among check nodes. 
The vertices are indexed by \(\mathcal{V} = \{1, 2, \dots, M\}\) (one for each check node). The edges are defined as \((m_1, m_2) \in \mathcal{E}\) if checks \(m_1\) and \(m_2\) share a variable, i.e., there exist a \( j\) such that \(H_{m_1,j} = H_{m_2,j} = 1\). Mathematically, the adjacency matrix is

\begin{equation}
    \mathbf{A} = \mathbf{H} \mathbf{H}^T - \operatorname{diag}(\mathbf{H} \mathbf{H}^T),
\end{equation}
where \(\mathbf{H} \mathbf{H}^T\) counts shared variables per pair, and the diagonal is zeroed to exclude self-loops. The adjacency matrix \(\mathbf{A}\) is used to determine conflicts for graph coloring: During greedy coloring, neighbors are identified as rows where \(A_{m,k} > 0\), to avoid assigning the same color.

Graph coloring assigns colors $c: \mathcal{V} \to \{1,2,\dots,\chi\}$, where $\chi$ is the chromatic number (i.e., the minimum number of required colors), such that no two adjacent vertices share the same color: $c(m_1) \neq c(m_2)$ if $(m_1, m_2) \in \mathcal{E}$. We employ a greedy coloring strategy. The check nodes are ordered as $m = 1,\dots,M$. For each check node $m$, we determine the set of colors already assigned to its neighbors:
\[
    U_m = \{c(k) \mid (m,k) \in \mathcal{E},\ c(k)\ \text{assigned}\},
\]
and then assign $c(m) = \min \{k \in \mathbb{N} \mid k \notin U_m\}.$
Based on the assigned colors, we define the layers as $L_k = \{m \mid c(m) = k\}$ for $k=1,\dots,\chi$. Since no two nodes within a layer are connected, each layer represents a non-conflicting group. In the layered scheduling framework, these layers are processed sequentially in each decoding iteration, and variable nodes are updated immediately after each layer is processed, enabling faster convergence and improved decoding efficiency~\cite{raeisi2021edge}.


\subsection{LNMS Message Passing Algorithm}
The LNMS algorithm is an iterative message-passing decoder that combines layered scheduling with the NMS approximation. Decoding proceeds over a maximum of \( \textit{max\_iter} \) iterations, indexed by \( t = 1, 2, \dots, \textit{max\_iter} \). In each iteration, the layers \( L_k \) for \( k = 1, \dots, \chi \), determined by the conflict graph coloring, are processed sequentially. Let \(L(c_i)\) denote the received channel LLR corresponding to bit \(i\). 
At each iteration \(t\), the variable-to-check node LLR from variable node \(i\) to check node \(j\) is denoted by \(L\!\left(q_{i \to j}^{(t)}\right)\), while the corresponding check-to-variable node LLR from check node \(j\) to variable node \(i\) is denoted by \(L\!\left(r_{i \leftarrow j}^{(t)}\right)\). 
These LLRs are updated iteratively according to the following steps.

\noindent\textbf{Step 1.}
The algorithm begins with standard initialization which we include here for the sake of completeness.
The code bit \(c_i \in \{0, 1\}\) is mapped to a binary phase-shift keying (BPSK) symbol \(x_i = 1 - 2c_i\). 
The received signal is \(y_i = x_i + n_i\), where \(n_i\) is additive white Gaussian noise with zero mean and variance \(\sigma^2\). 
The initial LLRs are $L(q_i^{(0)}) = L(c_i) = \frac{2 y_i}{\sigma^2}$, for all variable nodes \(i = 1, \dots, N\). 
All check-to-variable messages are initialized to zero as 
\(L\!\left(r_{i \leftarrow j}^{(0)}\right) = 0\) 
for all edges connecting check node \(j\) and variable node \(i\).
The variable \(q\) denotes the LLR associated with a variable node, whereas the variable \(r\) denotes the LLR associated with a check node.

\noindent\textbf{Step 2.} For each layer \( L_k \), the check nodes in \( L_k \) are processed in parallel. For a given check node \( j \in L_k\), let \( \mathcal{V}_j = \{i : H_{j,i} = 1\} \) be the set of connected variable nodes, with degree \( d_j = |\mathcal{V}_j| \). The variable-to-check messages are computed as
\begin{equation}
    L\!\left(q_{i \to j}^{(t)}\right)   = L(q_i^{(t-1)}) -  L\!\left(r_{i \leftarrow j}^{(t-1)}\right), \quad i \in \mathcal{V}_j.
    \label{eq:variable_to_check_message_update}
\end{equation}

\noindent\textbf{Step 3.} Next, the check-to-variable messages are approximated using the NMS rule. 
The updated check-to-variable message is given by
\begin{equation}
    L(r_{i \leftarrow j}^{(t)}) = \alpha \times \prod_{i' \in \mathcal{V}_j \setminus i} \operatorname{sgn}(L(q_{i' \to j}^{(t)})) \cdot \left(\min_{\neq i}\right),
    \label{eq:check_to_variable_message_update}
\end{equation}
where \(\alpha < 1\) is a normalization factor to compensate for the min-sum overestimation, $\mathcal{V}_j \setminus i$ represents \(\mathcal{V}_j\) excluding variable node \(i\), and \(\min_{\neq i}\) denotes the minimum magnitude excluding the message from each neighbor \(i\).

The vector of updated check-to-variable messages for check node \( j \) at iteration \( t \) is defined as \( \mathbf{L(r_j^{(t)})} = \{L(r_{i \leftarrow j}^{(t)}) : i \in \mathcal{V}_j\} \). The delta updates are computed as
\begin{equation}
    \Delta = \mathbf{L(r_j^{(t)})} - \mathbf{L(r_j^{(t-1)})}.
\end{equation}

The a posteriori LLRs are immediately updated:
\begin{equation}
    L(q_i^{(t)}) = L(q_i^{(t-1)}) + \Delta_i, \quad i \in \mathcal{V}_j.
    \label{eq:LLR_update}
\end{equation}
This immediate update after each layer accelerates information propagation compared to flooding scheduling. Steps 2 and 3 are repeated for all layers in an iteration, after which the algorithm proceeds to Step 4.

\noindent\textbf{Step 4.} 
After processing all layers in the iteration, hard decisions are made:
\begin{equation}
    \hat{c}_k = \mathbb{I}(L(q_k^{(t)}) < 0),
    \label{eq:decision}
\end{equation}
where \( \mathbb{I} \) is the indicator function (1 if true, 0 otherwise).

The syndrome is computed as
\begin{equation}
    \mathbf{s} = \mathbf{H} \mathbf{\hat{c}}^T.
    \label{eq:syndrome_computation}
\end{equation}

If \( \mathbf{s}\ = \mathbf{0} \), decoding succeeds, and the algorithm terminates early. Otherwise, the next iteration begins by repeating Steps 2 and 3 for all layers, followed by Step 4. This early termination reduces unnecessary computations \cite{jayawickrama2022improved}.

\subsection{SGBF Method}
If the LNMS algorithm fails to produce a zero syndrome after the maximum iterations, the SGBF method is activated as a post-processing step to correct residual errors.

The SGBF method identifies unreliable bits using a reliability metric that combines a posteriori LLR magnitudes with syndrome information. First, the per-bit failure count is computed:
\begin{equation}
    e_i = \sum_{j: H_{j,i}=1} s_j
    \label{eq:failure_count}
\end{equation}
where \(s_j\) is the \(j\)-th element of the syndrome \(\mathbf{s}\). The quantity $e_i$ represents the number of unsatisfied check equations involving the variable node \( i \). 

Our proposed reliability measure of the $i$-th bit is then defined as
\begin{equation}
    rel_i := \frac{|L(q_i^{(\textit{max\_iter})})|}{1 + \max(e_i, 1)}.
    \label{eq:reliability_measure}
\end{equation}
Roughly speaking, this proposed measure penalizes bits with low LLR confidence or high involvement in unsatisfied checks while the use of \( \max(e_i, 1) \) ensures a minimum penalty. 

The top \( T \) unreliable bits (i.e., those with the lowest reliability metrics) are selected: \( \{i_1, \dots, i_T\} \), where \( T \) is the bit-flipping set size \cite{chang2010check}. For each \( t = 1 \) to \( T \), a candidate channel LLR vector \(\mathbf{L(c)}_{t}\) is generated by flipping the sign of the LLR at position \( i_t \):

\begin{equation}
    \mathbf{L(c)}_{t} =
    \begin{cases}
        -L(c_i), & i = i_{t}, \\
        L(c_i), & i \neq i_{t}.

    \end{cases}
    \label{eq:sign_flip}
\end{equation}
This flip emulates correcting a potential bit error at that position.

Each candidate \( \mathbf{L(c)}_{t} \) is re-decoded using the LNMS algorithm with the same \textit{max\_iter}, producing updated a posteriori LLRs \( \mathbf{L(q)}_{t} \), hard decisions \( \mathbf{\hat{c}}_{t} \), and syndrome \( \mathbf{s}_{t} \).

The syndrome weight for each candidate is computed:
\begin{equation}
    w_{t} = \sum_{i=1}^{N-K} {s_t}_i. 
    \label{eq:syndrome_weight}
\end{equation}
The candidate with the minimum weight is selected: $
    {t}^* = \arg\min_{t} w_{t}.$
If \( w_{{t}^*}=0 \), a valid codeword has been found, and the final output is set to \( \mathbf{L(q)} = \mathbf{L(q)}_{{t}^*} \) and \( \mathbf{\hat{c}} = \mathbf{\hat{c}}_{{t}^*} \). Otherwise, the original LNMS output is retained.

This approach, with size \( T \), uses single-bit flips guided by the syndrome-weighted reliability metric to enhance error correction without excessive computational overhead. 

The overall proposed decoding algorithm is summarized in Algorithm \ref{alg:proposed-decoder}. The inputs are the received LLRs $\mathbf{L(c)}$, the parity-check matrix $\mathbf{H}$, the maximum number of iterations \textit{max\_iter}, and the bit-flipping set size $T$. The output is the estimated codeword $\mathbf{\hat{c}}$.

Graph coloring is performed on the conflict graph to determine the layers in line 1. Initialization is carried out in line 2. The LNMS iterations are implemented in lines 3--15: for each iteration, the layers are processed sequentially in lines 4--9, with variable-to-check messages, check-to-variable messages, and a posteriori LLR updates computed in parallel for check nodes within each layer in lines 6--8. After all layers, hard decisions are made in line 11 and the syndrome is computed in line 12. If the syndrome is zero in line 13, the estimated codeword is output and decoding stops in line 14.

If decoding fails after the maximum iterations, the SGBF phase is activated in lines 17--31: per-bit failure counts and reliabilities are computed in lines 18--19, the top \(T\) unreliable bits are selected in line 20, candidate LLR vectors are generated by flipping the least reliable bits and re-decoded using LNMS in lines 21--23, syndrome weights are computed in line 24, the candidate with the minimum syndrome weight is selected in line 26, and if it yields a zero syndrome in line 27, the corresponding a posteriori LLRs and codeword are adopted in line 28; otherwise, the original outputs are retained in line 30. The estimated codeword is output in line 33.

\algnewcommand\Input{\Statex \textbf{Input:} }
\algnewcommand\Output{\Statex \textbf{Output:} }

\begin{algorithm}[t]
\caption{The proposed decoder}
\label{alg:proposed-decoder}
\begin{algorithmic}[1] 
\Input{the received LLRs $\mathbf{L(c)}$, parity-check matrix $\mathbf{H}$, maximum iterations \textit{max\_iter}, bit-flipping set size $T$}
\Output{the estimated codeword $\mathbf{\hat{c}}$}
\State Perform graph coloring on the conflict graph $\mathcal{G}$ to determine the layers $L_1, \dots, L_\chi$.
\State Initialization: Set $L(q_i^{(0)}) = L(c_i)$ for all $i = 1, \dots, N$, and $L(r_{i\leftarrow j}^{(0)}) = 0$ for all edges.
\For{$\textit{iter} = 1 : \textit{max\_iter}$}
  \For{$k = 1 : \chi$}
    \For{each check node $j \in L_k$ \textbf{in parallel}}
      \State Compute variable-to-check messages using \eqref{eq:variable_to_check_message_update}
      \State Compute check-to-variable messages using \eqref{eq:check_to_variable_message_update}
      \State Update the a posteriori LLRs using \eqref{eq:LLR_update}
    \EndFor
  \EndFor
  \State Make hard decisions using \eqref{eq:decision}
  \State Compute the syndrome using \eqref{eq:syndrome_computation}
  \If{$\mathbf{s} = \mathbf{0}$}
    \State Output $\mathbf{\hat{c}}$ and stop.
  \EndIf
\EndFor
\If{decoding failed (non-zero syndrome)}
  \State Compute per-bit failure counts $e_i$ using \eqref{eq:failure_count} for all $i$.
  \State Compute reliabilities $rel_i$ using \eqref{eq:reliability_measure} for all $i$.
  \State The top \(T\) unreliable bits are selected: \( \{i_1, \dots, i_T\} \). 
  \For{$t = 1 : T$}
    \State Create candidate $\mathbf{L(c)}_{t}$ by flipping $L(c_{i_{t}})$ using \eqref{eq:sign_flip}.
    \State Re-run LNMS decoding on $\mathbf{L(c)}_{t}$ to obtain $\mathbf{L(q)}_{t}$, $\mathbf{\hat{c}}_{t}$, and $\mathbf{s}_{t}$.
    \State Compute the syndrome weight using \eqref{eq:syndrome_weight}.
  \EndFor
  \State Find ${t}^* = \arg\min_{t} w_{t}$.
  \If{$w_{{t}^*} = 0$}
    \State Set final $\mathbf{L(q)} = \mathbf{L(q)}_{{t}}^*$ and $\mathbf{\hat{c}} = \mathbf{\hat{c}}_{{t}}^*$.
  \Else
    \State Retain original $\mathbf{L(q)}$ and $\mathbf{\hat{c}}$.
  \EndIf
\EndIf
\State Output $\mathbf{\hat{c}}$.
\end{algorithmic}
\end{algorithm}

\section{Numerical Results}

In this section, we present simulation results obtained over the BI-AWGN channel using BPSK modulation.
For polar codes, we adopt the scaled min-sum decoding algorithm, using a scaling factor of 
$0.9375$ as suggested in \cite{yuan2014early}. The frame error rate (FER) is considered as the primary measure of error-correction performance.
The parity-check matrices are obtained using Algorithm~2 from~\cite{moradi2025high}.
The 5G-LDPC codes are implemented using the MATLAB 5G Toolbox, with a scaling factor of \(0.75\) and the NMS decoding algorithm.
The normalization factor \(\alpha\) used in our proposed decoding algorithm is also set to \(0.75\).

\subsection{Performance Comparison with Standalone LNMS}

\begin{figure}[t] 
  \centering
  \includegraphics[width=1\linewidth]{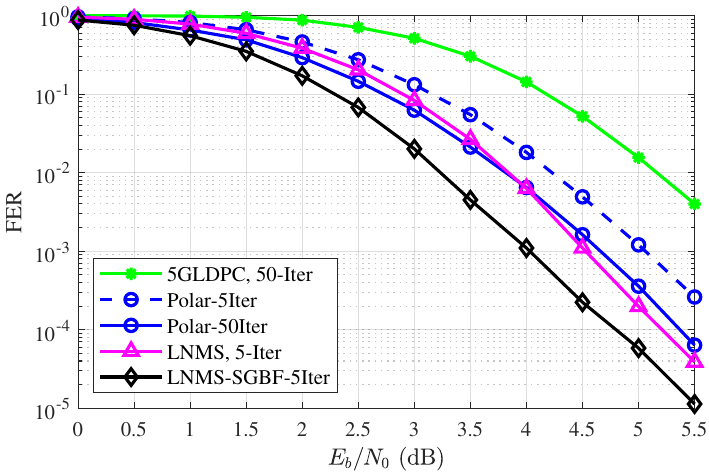}
  \caption{FER performance of the FDPC$(128,80)$ codes with LNMS decoding and LNMS with SGBF list correction, compared with the corresponding 5G-LDPC and polar codes.}
  \label{fig:fdpc128}
\end{figure}

Fig.~\ref{fig:fdpc128} shows the error-correction performance of our LNMS decoding algorithm in terms of FER versus energy per bit to noise power $E_b/N_0$, and compares it with the proposed improved LNMS incorporating SGBF (with \(T = 128\)) for the FDPC$(128,80)$ code under 5 iterations.
As shown in the plots, the use of SGBF provides nearly a \(0.5\)~dB coding gain.

\begin{figure}[t] 
  \centering
  \includegraphics[width=1\linewidth]{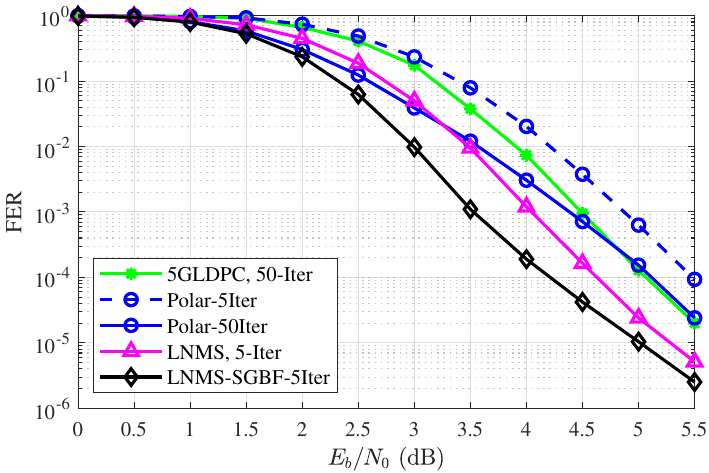}
  \caption{FER performance of the FDPC$(256,164)$ codes with LNMS decoding and LNMS with SGBF list correction, compared with the corresponding 5G-LDPC and polar codes.}
  \label{fig:fdpc256164}
  \vspace{-3mm}
\end{figure}

\begin{figure}[t] 
  \centering
  \includegraphics[width=1\linewidth]{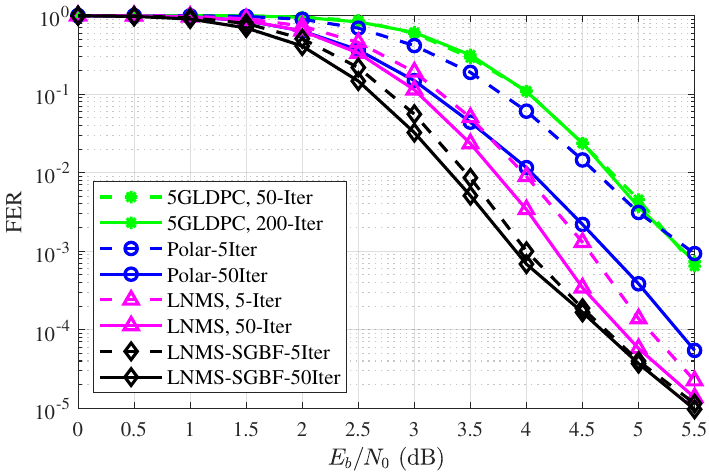}
  \caption{FER performance of the FDPC$(256,192)$ codes with LNMS decoding and LNMS with SGBF list correction, compared with the corresponding 5G-LDPC and polar codes.}
  \label{fig:fdpc256192}
\end{figure}

We have repeated similar experiments at other block lengths and rates. In particular, Fig.~\ref{fig:fdpc256164} compares the error-correction performance of our LNMS decoding algorithm with the proposed improved LNMS incorporating SGBF (with \(T = 128\)) for the FDPC$(256,164)$ code. Furthermore, Fig.~\ref{fig:fdpc256192} compares the error-correction performance of our LNMS decoding algorithm with the proposed improved LNMS incorporating SGBF (with \(T = 128\)) for the FDPC$(256,192)$ code. 

\begin{figure}[t] 
  \centering
  \includegraphics[width=1\linewidth]{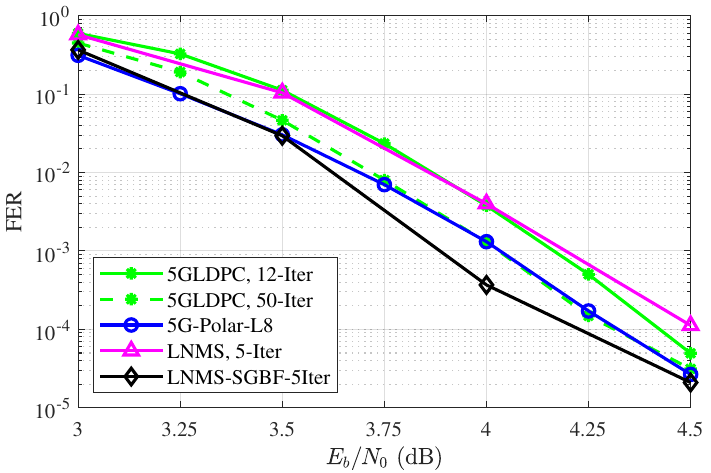}
  \caption{FER performance of the FDPC$(1024,844)$ codes with LNMS decoding and LNMS with SGBF list correction, compared with the corresponding 5G-LDPC and polar codes.}
  \label{fig:fdpc1024}
  \vspace{-3.5mm}
\end{figure}

The above scenarios demonstrate the advantages of the proposed decoder in the short block length regime (this regime is typically considered to be for block lengths up to a few hundreds). We have also conducted a similar experiment for a medium block length scenario (typically a few hundred to a few thousand bits). In particular, Fig.~\ref{fig:fdpc1024} compares the error-correction performance of our LNMS decoding algorithm with the proposed improved LNMS incorporating SGBF (with \(T = 128\)) for the FDPC$(1024,844)$ code. The number of layers in the proposed LNMS decoder is set to $4$ in all of our experiments. 

For reference, in all plots, we also include results for the NMS decoding of 5G-LDPC codes and the normalized BP decoding of polar codes, both with the same code length and data length. For the length $1024$, we consider polar list decoding with list size $8$. The results show consistent gains of FDPC codes with our proposed decoder compared to these other state-of-the-art codes.


\subsection{Impact of Bit-Flipping Set Size on Decoding Performance}

\begin{figure}[t]
  \centering
  \includegraphics[width=1\linewidth]{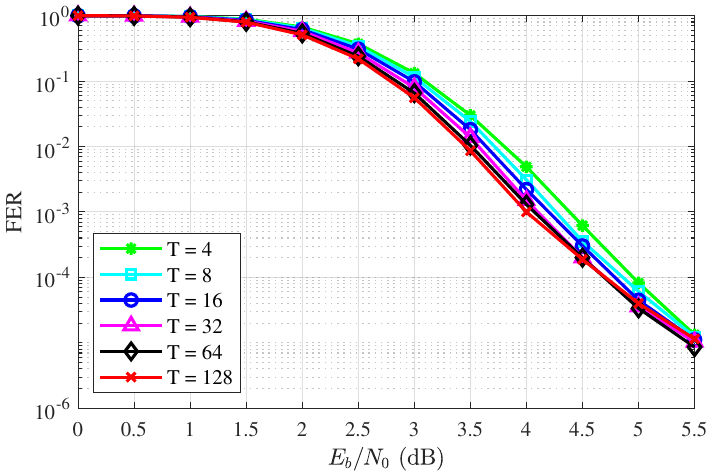}
  \caption{FER performance of the FDPC$(256,192)$ codes with LNMS with SGBF list correction for different bit-flipping set sizes.}
  \label{fig:list_size}
  \vspace{-5mm}
\end{figure}

We further investigate the impact of the flipping set size \(T\) in the SGBF phase on decoding performance. 
Simulations are conducted for the FDPC$(256,192)$ code with $5$ iterations. 
Fig.~\ref{fig:list_size} shows the FER performance for \(T = 4, 8, 16, 32, 64,\) and \(128\). 
As \(T\) increases, the FER performance improves, with larger values of $T$ providing stronger error-correction capability by considering more candidate bit flips in the post-processing phase. 
In particular, the improvement in the coding gain is more significant for smaller increases in \(T\) (e.g., from 4 to 32), but diminishes for larger values (e.g., from 64 to 128), indicating a trade-off between performance and computational complexity, which scales linearly with \(T\). 
Hence, \(T = 128\) represents a practical choice that achieves notable gains while maintaining manageable complexity.

\subsection{Decoding Complexity}

The complexity of the proposed LNMS decoder is comparable to that of the NMS decoder for 5G-LDPC codes, with per-iteration complexity proportional to the number of edges in the Tanner graph, similar for both FDPC and 5G-LDPC codes at the considered parameters. Layered scheduling in LNMS maintains computational complexity while accelerating convergence and potentially reducing average iterations via early termination.

The SGBF method adds complexity only upon LNMS failure, requiring $T$ additional LNMS runs. The average increase is $P_{\text{fail}} \times T \times C_{\text{LNMS}}$, where $P_{\text{fail}}$ is the LNMS failure probability which is equivalent to its FER and $C_{\text{LNMS}}$ is the complexity of one LNMS run. Failures are rare at high SNR, making the average additional decoding complexity negligible in high-reliability scenarios. However, for fair comparison, we adjust 5G-LDPC iterations to match the proposed decoder's average complexity. Simulations show the FDPC code with the new decoder outperforms 5G-LDPC codes by 0.5--1.0\,dB at FER$=10^{-3}$ under these conditions, highlighting its superiority.

\section{Conclusions}
In this paper, we proposed an LNMS message-passing decoding algorithm and an SGBF list correction method to enhance the error-correction performance of FDPC codes. By employing conflict graph coloring for layered scheduling, the proposed decoder achieves faster convergence compared to traditional flooding schedules. The SGBF method effectively addresses decoding failures by flipping the least reliable bits based on a syndrome-weighted reliability metric, generating a set of candidates and selecting the ones with the minimum syndrome weights. Simulation results demonstrate that the proposed decoder outperforms the standalone LNMS decoder. The numerical results highlight the effectiveness of both our proposed decoder and the bit-flipping method.

\bibliographystyle{IEEEtran}
\bibliography{bibliography}
\end{document}